\documentclass[aps,prd,amsmath,twocolumn]{revtex4}

\usepackage{amsmath}
\usepackage{bm}
\usepackage[dvips]{color,graphicx}
\usepackage{epsfig}
\usepackage{dcolumn}

\begin{document}

\title{$Q^2$-evolution of the electric and magnetic polarizabilities of the proton}

\author{A.~Sibirtsev and P.~G.~Blunden}
\affiliation{Department of Physics and Astronomy,
University of Manitoba, Winnipeg, MB, Canada\ \ R3T~2N2}

\begin{abstract}
The generalized Baldin sum rule at finite four-momentum transfer $Q^2$
is evaluated utilizing a structure function 
parameterization fit to recent experimental data.
The most recent measurements on 
$F_1$ from Hall C at Jlab, as well as the $F_2$ structure function data from
Hall B at Jlab and SLAC, were used in constructing our parameterization.
We find that at $Q^2$ below 1~GeV$^2$ the dominant contribution to
the electric and magnetic polarizabilities of the nucleon comes from the
resonance region.
\end{abstract}

\maketitle
\section{Introduction}

Understanding the internal structure of strongly 
interacting particles is one of the major goals of low-energy QCD. 
The study of the response of baryons to an external electromagnetic field 
via the multipole excitation mechanism provides direct access to the 
internal structure.

The parameters to describe that response 
are electric, magnetic, and spin-dependent polarizabilities. The 
polarizability is an elementary structure constant
that is related to the deformation and stiffness of the baryon.
Furthermore, the physical content of the polarizabilities is an 
effective multipole interaction for the coupling of the electric 
and magnetic fields of the photon with the
internal structure of the baryon~\cite{Babusci1, Holstein}.

Evaluation of the electric $(\alpha)$ and magnetic $(\beta)$ 
polarizabilities of the nucleon has attracted much attention, 
both phenomenological and theoretical. The 
$Q^2$-dependence of the polarizabilities gives information on
the polarization density in the nucleon. The $Q^2$-evolution of the sum 
of both polarizabilities can be determined through the generalized 
Baldin sum rule~\cite{Drechsel}, namely
\begin{equation}
\label{esr}
{\ \alpha}(Q^2)+{\ \beta}(Q^2) =\frac{8 \, \alpha_{\rm em} \, M}{Q^4}
\int_0^{x_\pi}\!x\, F_1(x,Q^2)\,dx,
\end{equation}
where $Q^2$ is four-momentum transfer squared, $\alpha_{\rm em}$ is the fine structure
constant, $M$ stands for the nucleon mass, and $F_1$ is the nucleon structure 
function. Here $x$ is the Bjorken scaling variable,
\begin{equation}
x=\frac{Q^2}{W^2-M^2+Q^2},
\end{equation}
where $W$ is the invariant mass of the final state, and $x_\pi$ 
corresponds to pion threshold.
This implies that $F_1$ should be taken up to an infinite 
energy. Indeed it is necessary to use 
quite reliable models for the $W$-dependence as well as $Q^2$-dependence of the
structure function in order to reduce the
uncertainty of the sum rule evaluation. 

Equation~(\ref{esr}) provides information about the $Q^2$-evolution 
of the polarizabilities. An evaluation of the 
$Q^2$-dependence of the sum of electric and magnetic polarizabilities was 
done in 2006 by Liang {\em et al.}~\cite{Liang1}. There was not much 
data on the $F_1$ structure function in the relevant kinematic region
at that time.
At  high energies the SLAC Rosenbluth 
data~\cite{Whitlow} were used. In the resonance region the E94-110 
measurements~\cite{Liang2,Liang3} in Hall C at Jefferson 
Laboratory (JLab) from 2004 were used.

Up to now this evaluation of the $Q^2$-dependence of polarizabilities stands as the
present ``state of the art''. However, in 2013 an updated version of the E94-110 measurement 
appeared~\cite{Liang4}. This has motivated us to reexamine the generalized
Baldin sum rule.

At $Q^2=0$ the sum of electric and magnetic 
polarizabilities of the nucleon can be related to the unpolarized photo-absorption
cross section $\sigma_{\gamma N\to X}$ as 
\begin{equation}
\label{abs}
{ \alpha}+{\beta}=\frac{1}{2\pi^2}
\int_{\nu_\pi}^{\infty}\frac{\sigma_{\gamma N\to X}(\nu )}{\nu^2}d\nu,
\end{equation}
where $\nu$ is the photon energy and $\nu_\pi$ is pion photo-production 
threshold. It is clear that at $Q^2=0$ Eq.~(\ref{esr}) converges to Eq.~(\ref{abs}),
since
\begin{equation}
x=\frac{Q^2}{2M\nu}.
\end{equation}
Indeed, Eq.~(\ref{abs}) is the original formulation of the Baldin sum rule~\cite{Baldin,Lapidus}.

The evaluation of the Baldin sum rule requires knowledge of the 
energy-dependence of $\sigma_{\gamma N\to X}(\nu)$ up to $\nu \to\infty$. 
However, as the integral is weighted by $1/\nu^2$, the contribution at low energies dominates the integral.

The sum of the polarizabilites for the proton was
evaluated by Damashek and Gilman~\cite{Damashek} in 1970, and amounts to
\begin{equation}
{\alpha}+{\beta} = (14.2 \pm 0.3)\times 10^{-4} \,{\rm fm^3}.
\end{equation}
A more recent calculation of Eq.~(\ref{abs}) was done by 
Babusci {\em et al.}~\cite{Babusci}, with the result
\begin{equation}
{\ \alpha}+{\ \beta} = (13.69 \pm 0.14)\times 10^{-4} \,{\rm fm^3}.
\end{equation}
The current PDG~\cite{PDG} averaged experimental values for 
electric and magnetic polarizabilities for proton are
\begin{subequations}
\begin{eqnarray}
\alpha&=& (12.0 \pm 0.6)\times 10^{-4}\,{\rm fm^3},\\
\beta &=& (1.9 \pm 0.5)\times 10^{-4} \,{\rm fm^3}.
\end{eqnarray}
\end{subequations}
Finally, a Chiral Perturbation Theory calculation~\cite{Bernard} predicts
\begin{equation}
{\alpha}+{\beta} = (14.0 \pm 4.1)\times 10^{-4} \,{\rm fm^3}
\end{equation}
for the proton. An evaluation of the generalized polarizabilities
at low $Q^2$ was considered by Hemmert {\em et al.}~\cite{Hemmert}.

A review of nucleon polarizabilities at $Q^2=0$ extracted from 
experimental data, as well as given by theoretical calculation, can be 
found in Ref.~\cite{Babusci1}. A more recent review within the effective
field theory approach can be found in Ref.~\cite{Griesshammer}.
The polarizabilities at $Q^2=0$ provide a 
lower limit for the $Q^2$-dependence of the generalized Baldin sum rule.

Here we calculate the generalized Baldin sum rule, aiming to obtain the 
$Q^2$-dependence of the sum of generalized electric and magnetic 
proton polarizabilities. As a corollary, we obtain a convenient parameterization
of the $F_1$ structure function from recent experimental data which is
valid in the low-$Q^2$, low $W$ kinematic region.

The most recent results~\cite{Liang4} for the $F_1$ structure function
measured at Jlab Hall C were used for the contribution in 
the resonance region. Furthermore, we adopted the results on the $F_2$ 
structure function in the resonance region obtained at JLab.
For that we apply the relation between $F_2$ and $F_1$ given by the
ratio of longitudinal to transverse cross sections. That procedure 
provides us reasonable confidence in constructing the parameterization 
of the structure functions. 

At energies above the resonance region we 
adopt the Regge approach. It is shown here that the Regge results are in good 
agreement with MRST leading twist fit~\cite{MRST,MSTW} at 
$Q^2>1$~GeV$^2$. Note that the MRST parton distribution function is not
provided for $Q^2 < 1$~GeV$^2$.
 
The paper is organized as follows. In Sec.~II we show the details of 
our parameterization of the structure functions, and compare it
with experimental results. The evaluation of the generalized
Baldin sum rules is given in Sec.~III. The paper ends with a summary.

\section{The parameterization}

For the $F_1$ structure function we use the parameterization developed 
in Ref.~\cite{Sibirtsev}. The parameters of that model were
obtained from the fit of experimental results for
the $F_2$ structure function 
measured at JLab~\cite{Liang3,Osipenko,Niculascu,Malace}, 
as well as obtained at SLAC~\cite{Whitlow1}.
The structure functions $F_1$ and $F_2$ are related through the ratio of the
longitudinal $\sigma_L$ to transverse $\sigma_T$ virtual photon 
cross sections as
\begin{equation}
R =\frac{\sigma_L}{\sigma_T}=\frac{F_2}{2xF_1} 
\left[ 1+\frac{4M^2 x^2}{Q^2}\right]-1.
\end{equation}

\begin{figure}[t]
\vspace*{-1mm}
\hspace*{1.5mm}\includegraphics[width=9.3cm]{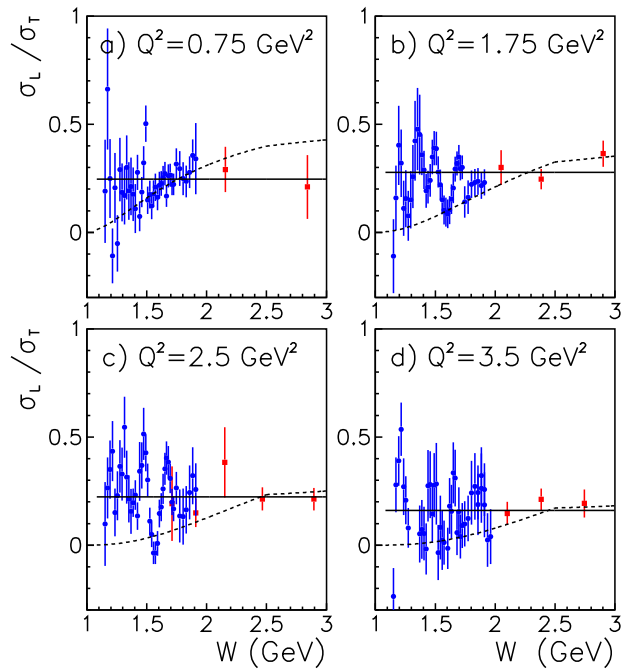}
\vspace{-7mm}
\caption{(Color online). The ratio of the longitudinal $\sigma_L$ to transverse 
$\sigma_T$ virtual photon cross sections as a function of energy $W$ shown 
for different $Q^2$. The circles are the results~\cite{Liang4} 
from Jlab Hall~C. The squares are the the results from SLAC~\cite{Whitlow}.
Solid lines are our parameterization given by Eq.~(\ref{our}). Dashed lines 
show the parameterization from Ref.~\cite{Ricco}.}
\label{ratio}
\end{figure}

Figure~\ref{ratio} shows the experimental results for the ratio. The circles 
indicate the data~\cite{Liang4} from Jlab Hall C, while the squares 
are the data from SLAC~\cite{Whitlow}. The ratio is shown for 
different $Q^2$. Here solid lines are our parameterization, given as
\begin{equation}
R = 0.014\, Q^2 \, \left [\exp(-0.07\, Q^2 ) + 41\, \exp(-0.8\, Q^2 )
\right],
\label{our}
\end{equation}
with $Q^2$ in units of GeV$^2$.

Actually the parameters of Eq.~(\ref{our}) were obtained from a fit 
of the experimental results shown in Fig.~\ref{ratio}. Since the data at 
$W<2$~GeV are given with quite large statistical and experimental errors,
we were not able to fit the fine structure of the ratio in the resonance 
region. Equation~(\ref{our}) provides a correct limit at $Q^2=0$ ({\em i.e.}
$R=0$), since at the real photon point $\sigma_L=0$. Equation~(\ref{our}) was
used to obtain  the parameterization for the $F_1$ structure function from that 
given~\cite{Sibirtsev} for $F_2$.

Now we give the details of the $F_2$ parameterization, since these 
were not published in Ref.~\cite{Sibirtsev}. We did consider two 
regions with respect to energy $W$, namely the resonance and 
DIS regions.

At $W<1.9$~GeV we use the isobar model following the analyses of 
Refs.~\cite{Cone,Stein,Brasse}. The contributions from 
four resonances, namely $P33 (1232)$, $D13 (1520)$, $F15 (1680)$ and 
$F37 (1950)$, were considered. The resonance properties adopted in our
analysis are listed in Table~\ref{tab1}. Note that the masses and 
widths of the resonances were not taken from PDG~\cite{PDG}, but were
obtained from the fit of experimental results for the $F_2$ structure function.

\begin{table}[b]
\caption{Resonance parameters used in our isobar model.}
\centering
\begin{tabular}{c c c c c c c}
\hline 
Res. & $M_R$ (GeV) &  $\Gamma_R$ (GeV) & $l$ & $x_R$&
$b$ (GeV$^{-2})$ & $A_R$ \\ 
\hline
$P33$ & 1.22 &0.119 & 1 &  0.16& 1.51   &  723.1 \\
$D13$ & 1.52 & 0.127 & 2 &  0.35&1.32  & 214.6\\
$F15$ & 1.70 & 0.117 & 3 &  0.35& 0.81  &95.6 \\
$F37$ & 1.90 & 0.28 & 3 &  0.35& 1.11  & 68.2 \\ 
\hline
\end{tabular}
\label{tab1}
\end{table}

The resonance construction was parameterized by a relativistic Breit-Wigner shape as
\begin{equation}
\sigma_R=\frac{A_R \,M_R^2 \, \Gamma_1 \,\Gamma_2\, G(Q^2)}{(M_R^2-W^2)^2+M_R^2\Gamma_1^2}\,
\left( \frac{k_R}{k} \right)^2,
\end{equation} 
and we account for the energy-dependence of the width, namely
\begin{subequations}
\begin{eqnarray}
\Gamma_1=\Gamma_R\left(\frac{q}{q_R}\right)^{2l+1}\left(\frac{q^2+x_R^2}{q_R^2+x_R^2}
\right)^l, \\
\Gamma_2=\Gamma_R\left(\frac{q}{q_R}\right)^{2l}\left(\frac{k^2+x_R^2}{k_R^2+x_R^2}
\right)^l.
\end{eqnarray}
\end{subequations}
Furthermore
\begin{subequations}
\begin{eqnarray}
k=\frac{\lambda^{1/2}(W^2,m_N^2,Q^2)}{2W}, \\
k_R=\frac{\lambda^{1/2}(M_R^2,m_N^2,Q^2)}{2M_R}, \\
q=\frac{\lambda^{1/2}(W^2,m_N^2,m_\pi^2)}{2W}, \\
q_R=\frac{\lambda^{1/2}(M_R^2,m_N^2,m_\pi^2)}{2M_R}, 
\end{eqnarray}
\end{subequations}
with the kinematical function $\lambda$ defined as
\begin{equation}
\lambda(x,y,z)=(x-y-z)^2-4yz.
\end{equation}
We introduce form factors at the interaction vertices, parameterized by an 
overall exponential function as
\begin{equation}
G(Q^2)=\exp(-b Q^2),
\end{equation}
with cut off parameters $b$ listed in Table~1.
The parameters were fit to data~\cite{Sibirtsev}
on the $F_2$-structure function. 

With respect to Ref.~\cite{Christy}, we do not consider $S11(1535)$, $S11(1650)$,
and $P11(1440)$ resonances. We found that it is difficult to separate
these resonances from others in performing the fit and keep the resonance
properties ({\em e.g.} strength, width) as free parameters.
Since our strategy was to describe the experimental
results on the structure function, not to study baryons, this approach seems
quite reasonable. 

Moreover, in the resonance region we consider in addition the non-resonant background contribution
$F_2^{\rm bg}$. It was parameterized as
\begin{subequations}
\begin{eqnarray}
F_2^{\rm bg}&=& c_1 c_2 {(1-x)^{1.5} \over x^{0.6}},\\
c_1&=&4\,\pi^2\, \alpha_{\rm em}{Q^2+a_\nu^2\over Q^2 a_\nu^2 (1-x)}, \\
c_2&=&0.0037575 + 0.075834\, Q^2 \nonumber \\ && +\, 0.024600\, Q^4
-0.0099514\, Q^6, \\
x &=&{Q^2\over 2 (M+m_\pi) a_\nu},\\
a_\nu&=&{W^2+Q^2-(M+m_\pi)^2\over 2 (M+m_\pi)}.
\end{eqnarray}
\end{subequations}

\begin{figure}[t]
\vspace*{-1mm}
\hspace*{1mm}\includegraphics[width=9.3cm]{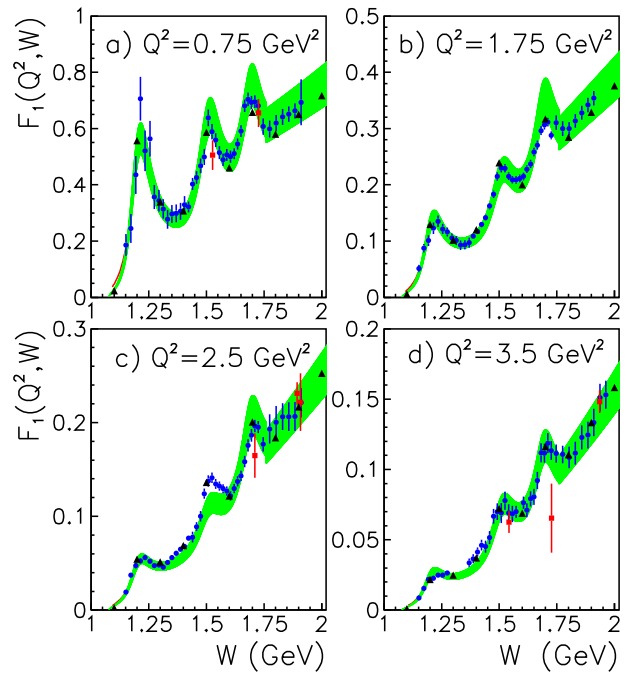}
\caption{(Color online). $F_1$ structure function in the resonance region ($W<2$~GeV)
shown for different $Q^2$ values. The circles are the 
results~\cite{Liang4} from JLab Hall~C, while the squares are 
from SLAC~\cite{Whitlow}. The shaded (green) band represents the 
uncertainty on our fit. The triangles are the results from the 
parameterization given in Ref.~\cite{Christy}.}
\label{fig:res}
\end{figure}

\begin{figure}[t]
\vspace*{-1mm}
\hspace*{1mm}\includegraphics[width=9.3cm]{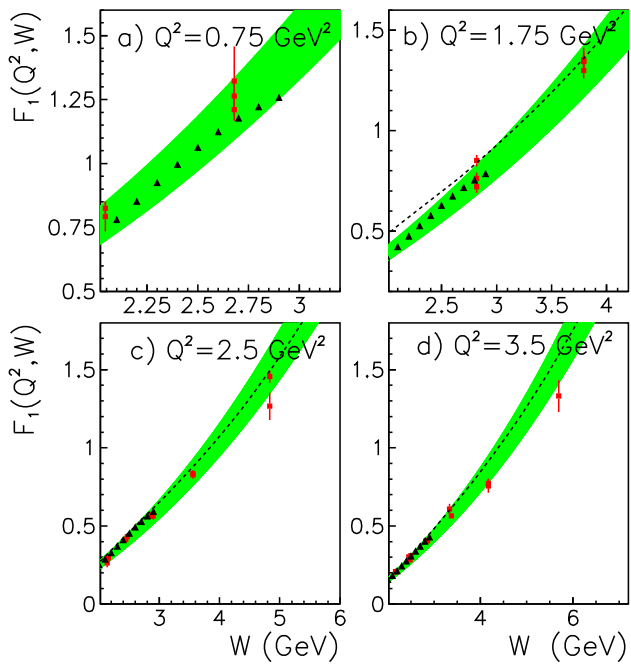}
\caption{(Color online). $F_1$ structure function above the resonance region shown for
different $Q^2$ values. The squares are 
from SLAC~\cite{Whitlow}. The shaded (green) band represents the 
uncertainty on our fit. The triangles are the results from the 
parameterization given in Ref.~\cite{Christy}. The dashed lines 
illustrate results from MRST~\cite{MRST,MSTW}.}
\label{fig:abres}
\end{figure}

Figure~\ref{fig:res} shows our calculations together with data
for the $F_1$ structure function. The circles are experimental results from
JLab~\cite{Liang4}, and the squares indicate data from SLAC~\cite{Whitlow}.
The shaded band represents the uncertainty in our fit. The triangles are
results from the parameterization in Ref.~\cite{Christy}. It is clear 
that both parameterizations describe the data quite reasonably, although they are
constructed in a different manner. 

Above the resonance region we adopt the Regge model~\cite{Capella,Kaidalov,Levin}.
The advantage of the Regge approach is that it can be used as well at low
$Q^2$, where the MRST PDF description cannot be applied. The comparison of Regge results with
MRST and data is given in Fig.~\ref{fig:abres}. There is a 
reasonable agreement between MRST and Regge results, as well as describing the
available data.

\section{Polarizabilities}

\begin{figure}[ht]
\vspace*{4mm}
\hspace*{1mm}\includegraphics[width=7.3cm]{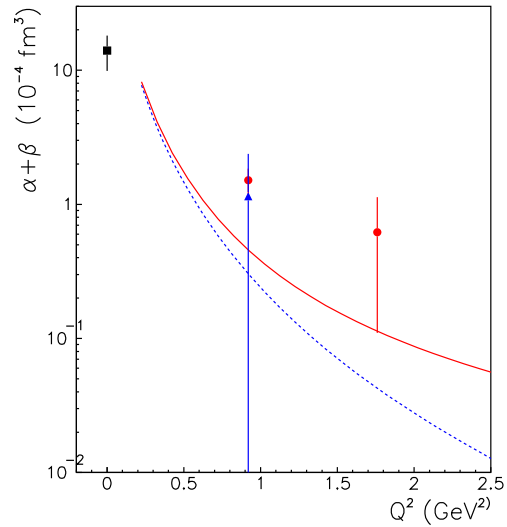}
\caption{(Color online). The sum of the electric and magnetic polarizabilites of the proton
as a function of $Q^2$.
The dashed line indicates the result for $W<2$~GeV, while the solid line 
show our calculation for the full range of energy. The square is the 
prediction~\cite{Bernard} given by Chiral Perturbation Theory at
$Q^2=0$. Circles are the results from Hall A obtained with a dispersion relation
approach and using data sets I-a and II~\cite{halla}. The triangle result is for data set 
I-b.}
\label{fig:pol}
\end{figure}

Now with the given parameterization for the $F_1$ structure function, we calculate the
sum of electric and magnetic polarizabilities of the proton. Since the parameters
of the model were fitted at $Q^2>0.225$~GeV$^2$, the results shown in 
Fig.~\ref{fig:pol} are for the relevant region. Here the dashed line indicates
the result for energies $W<2$~GeV. The solid line is the result for the
full range of energy. The square illustrates the prediction~\cite{Bernard} 
from ChPT.

We also show recent experimental results from Hall A given at $Q^2=0.92$~GeV$^2$ 
and 1.76~GeV$^2$ and evaluated by the dispersion relation approach. Unfortunately 
statistical and systematic errors of the data are too large, so it is not possible to make
solid conclusions on the compatibility of our analysis and measurements.
It appears that further precise experiments are required.
Furthermore, at $Q^2< 1$~GeV$^2$ the 
dominant contribution to polarizabilities comes from the resonance region. 

\section{Summary}

We evaluated the generalized Baldin sum rule using a structure function 
parameterization fit to experimental data. The most recent data on 
$F_1$ from Hall C at Jlab were used in our analysis. As well, we
utilized the $F_2$ structure function data collected by the Hall B Collaboration 
at JLab and the $F_2/F_1$ ratio. SLAC results are also included in our
analysis. We found that at $Q^2<1$~GeV$^2$ the dominant contribution 
to the sum of electric and magnetic polarizabilities comes from the
resonance region, {\it i.e.} at energies $W<2$~GeV. 

Further study is necessary at $Q^2<0.2$~GeV$^2$ to establish the transition to
the real photon point. In the absence of experimental results in that region we 
could not develop a phenomenological model. It is important to construct
a reliable parameterization for $F_1$ structure function if possible. We 
keep this study in progress. 

We note that linear extrapolation of our results to the real photon point does 
not allow us to get polarizability at $Q^2=0$. That issue as well requires a more
detailed study at low $Q^2$.

\begin{acknowledgments}
We would like to thank M.E.~Christy and P.E.~Bosted for providing us 
with their code, and V.~Tvaskis, E.~Epelbaum, and Ulf-G.~Mei{\ss}ner for many useful
discussions.
\end{acknowledgments}


\end{document}